Chapter 3

# Orchestrating the Implementation of the Smart City

## Abstract

This chapter examines the six smart city dimensions that serve as pillars in smart city projects. These dimensions are crucial in the development and evaluation of smart city initiatives, representing key areas for consideration. This chapter offers a detailed analysis of the smart city ecosystem, focusing on the governance, environment, people, living, mobility, and economy dimensions. It challenges the prevailing media portrayal of the smart city strategy and engages in the current academic debate surrounding these dimensions. This chapter defines, discusses, and explains each dimension, incorporating case studies from cities such as Copenhagen, San Francisco, Lisbon, and Barcelona. It also includes interviews and factual data to highlight the internal implementation and objectives of the smart city within each dimension. This chapter provides a comprehensive understanding of the smart city ecosystem, its implementation, and the potential benefits and challenges associated with each dimension.

## 3.1 Understanding Smart City Dimensions

In the previous chapter, we discussed the concept of smart cities, their definition, and the importance of embracing this approach for sustainable urban development. In this chapter, the aim is to delve deeper into the various dimensions of smart cities and how vital it is to orchestrate these dimensions harmoniously to touch upon all crucial aspects of life and urban dynamics.

As Dirks and Keeling (2009) emphasized, organic integration of a city's various systems (transportation, energy, education, health care, buildings, physical infrastructure, food, water, and public safety) is essential in creating a smart city. Moreover, Carvalho (2015) stresses that infusing intelligence into each subsystem of a city is not sufficient to develop a smart city. Instead, it should be treated as an organic whole.

In light of the complexity of managing the smart city concept in a holistic way, various researchers have attempted to break down the concept into smaller, more manageable features and dimensions. Giffinger, Fertner, Kramar, and Meijers (2007) is one of the first scholars to identify the six main components of a smart city: smart economy, smart mobility, smart environment, smart people, smart living, and smart governance. These components were later associated with different aspects of urban life and have become essential to recognize that these dimensions are interrelated and should be balanced to achieve the best outcomes for a smart city (Albino, Berardi, & Dangelico, 2015).

The six dimensions of a smart city are not isolated entities. Instead, they are interconnected, with each dimension influencing and being influenced by the others (Linde, Sjödin, Parida, & Wincent, 2021; Vanolo, 2014). The key to a successful smart city lies in the ability to orchestrate these dimensions effectively to create a cohesive urban ecosystem that addresses the challenges and opportunities of the 21st century.

One of the critical aspects of this orchestration is the role of information and communication technologies (ICTs) as we anticipated in Chapter 2. Nam and Pardo (2011a) argue that a truly smart city emerges when investments in human and social capital, along with ICT infrastructures, fuel sustainable growth and enhance the quality of life. While ICT is a vital component of smart cities, it is not the only factor. It is essential to recognize the importance of human capital and the necessity for smart governance that fosters creativity, diversity, and inclusivity. Regarding this matter, this perspective clearly emerges in the current trajectories of smart city development. For example, in an interview with a public manager, it emerged that:

> Obviously, the technology we have at our disposal as a city and the ability to operate on multiple fronts make us in line with the digital standards proposed at the country and European levels. The main problem is the management of this technology, which is why we need internal resources and high-skilled human capital to implement these practices internally (with the help of external suppliers). In this regard, the example of our municipality is perfect because we have established a dedicated office to bring together internal expertise and technology. We can say that until three years ago, technological innovation did not have a clear definition within a particular sector in our offices, but rather was a service, with a less structured dimension and half of the personnel we have today. However, we have now established a dedicated department that operates in this direction, and this has also been the winning card that, in retrospect, has allowed us to overcome the Covid period unscathed and initiate this digital transition, contributing to all six dimensions of the smart city.

Another crucial aspect of harmonizing the smart city dimensions is taking a human-centric approach. As Holland (2008) suggests, smarter cities begin with human capital rather than relying solely on ICT. By focusing on education and leadership and fostering an environment where entrepreneurship is accessible to all citizens, smart cities can ensure that the "smart people" dimension is fully realized (Neirotti, De Marco, Cagliano, Mangano, & Scorrano, 2014). This includes addressing barriers related to language, culture, education, and disability and encouraging creativity, flexibility, open-mindedness, and participation in public life (S. Y. Lee, Florida, & Gates, 2010; Marchesani, Masciarelli, & Quang, 2022).

Building on this discussion, in the upcoming sections of this chapter, we will delve deeper into each of the six smart city dimensions (see Fig. 2).

By examining these dimensions individually, we can better understand their role and potential in smart city projects. Through practical examples and insights from interviews with experts, we will highlight the importance of balancing these dimensions for a truly intelligent, sustainable urban environment.

As we continue our exploration of the various dimensions of smart cities, it is crucial to keep in mind the importance of harmoniously orchestrating these dimensions within the urban ecosystem. By doing so, we can create cities that not only leverage technology to improve efficiency and sustainability but also foster human capital, inclusivity, and creativity to enhance the overall quality of life for all residents.

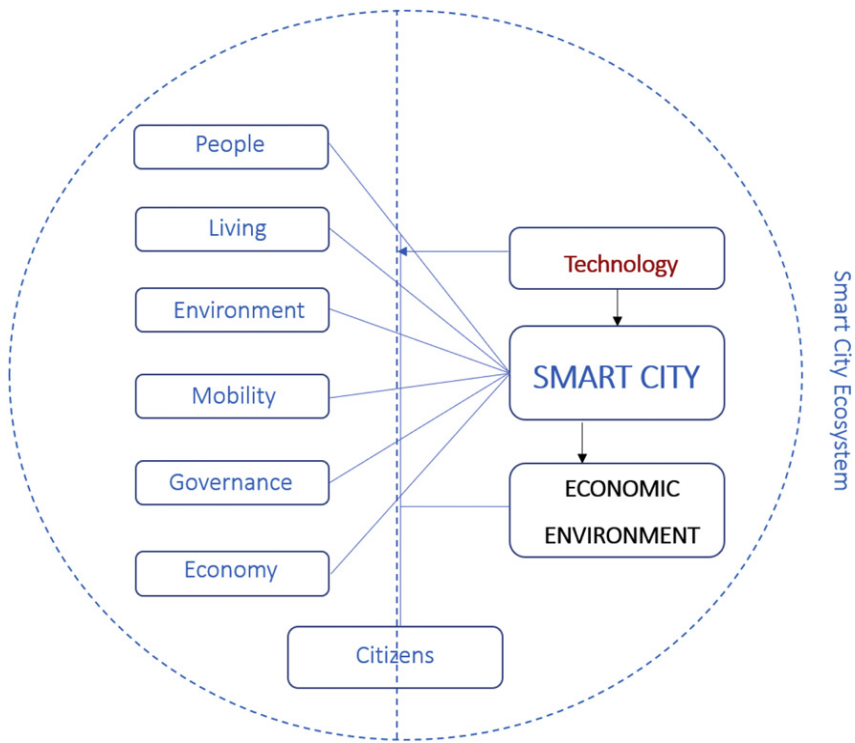

Fig. 2.   Six Dimensions in the Smart City Ecosystem. *Source*: Created by author.

### 3.1.1 Smart Governance

As previously discussed, we could say that the digital transformation is altering governance models in a disruptive way. Governance is the enabling environment that requires adequate legal frameworks and efficient processes to enable the responsiveness of government to the needs of citizens (UN Habitat, 2008). At the same time, it can also be defined as interaction and collaboration of different stakeholders in decision-making processes (Albino et al., 2015; Pittaway & Montazemi, 2020). The concept is commonly used to describe the action or manner of governing a state, an organization, or other constellation of actors. This shows that government and governance are related but different concepts.

Smart governance is defined as "*the capacity of employing intelligent and adaptive acts and activities of looking after and making decisions about the urban environment*" (Scholl & Alawadhi, 2016, p. 22), and it is "related to participation in decision-making processes, transparency of governance systems, availability of public services and quality of political strategies" (Vanolo, 2014, p. 887). According to these definitions, smart governance can be seen as the basis for smart, open, and participatory government. These concepts play a key role in the growing discourse on smart cities, so we may expect that ICTs play a key role in smart government as part of wider models of smart governance (S. Y. Tan & Taeihagh, 2020). And indeed, this concept emerges clearly also in interviews with smart city managers, who state that:

> In the development of smart city projects, governance and the way the city operates and interacts with its citizens is obviously fundamental component that ties different dimensions together and drives the development of smart cities. Without the strong push from the mayor and the administration, it would not have been possible to invest so many resources and efforts in smart practices in the city, which is now our strength in being one of the most digitally advanced cities in Italy. It is clear that these investments and efforts represent a great risk because the public may be more or less receptive, and we cannot be coercive in the use of digital services since the public is diverse and does not have homogeneous competencies. However, I can say that thanks to the vision of the current public administration and the creation of central offices and figures, the smart city project has had positive outcomes that are not only giving us great results but also driving us to push and invest even more in this direction.

Today, governments from all over the world are being challenged to become more innovative while at the same time reducing costs, operating in a connected environment, and engaging stakeholders in solving societal problems. In this context, the use of ICT in government has become a strategy for administrative reforms at all levels of government, and the pervasive use of ICTs in governance led to the creation of new governance perspectives such as e-government and digital government as a part of the smart governance dimension (Gagliardi et al., 2017; Pittaway & Montazemi, 2020; Sancino & Hudson, 2020). E-government or digital government can thus be seen as a transformational effort through the introduction of ICT in government organizations to achieve several objectives: improve the efficiency of public sector activities and service delivery but also transform government functions, internal organization, and practices through a citizen-centric approach; increase transparency; promote openness; and reduce corruption (Maria Lluïsa Marsal-Llacuna, Colomer-Llinàs, & Meléndez-Frigola, 2015; Maria Luisa Marsal-Llacuna & Segal, 2017; Neumann, Matt, Hitz-Gamper, Schmidthuber, & Stürmer, 2019; Shelton, Zook, & Wiig, 2015; Vu & Hartley, 2018). Therefore, governments around the world are adopting e-government strategies to improve the use of ICT. These emerging strategies mainly revolve around three added value "generators": administrative efficiency and interoperability, service improvement, and citizen centricity. In this sense, Tura and Ojanen (2022) argue that smart governance enables, with the help of ICT, the participation and collaboration of various actors in decision-making. For several authors (Abid, Marchesani, Ceci, & Masciarelli, 2023; Borrás & Edler, 2020; Gagliardi et al., 2017; Marchesani, Masciarelli, & Bikfalvi, 2023; Pinna, Masala, & Garau, 2017, 2022) this smart city governance makes use of new technologies and develops innovative governance combinations. Electronic platforms represent the use of ICTs to encourage citizen participation in decision-making, improving the provision of information and services, and enhancing transparency, accountability, and credibility (Mergel, Edelmann, & Haug, 2019). It is thus showing the possibility of a collaborative link between government and society, with ICT as an ally. After all, the main goal of smart city governance is not just the use of new technologies but the contribution of the urban environment that should focus on the community, network, and participants (Sancino & Hudson, 2020). In this sense, the pandemic has accelerated the relationship between governments, technology, and citizens, as reported in the interview:

> Technology is a useful means, and we have become fully aware of this during the pandemic. The only way to provide services to citizens was to become digital and push citizens to digitize not by choice but out of survival. In fact, many internal services that were already available, as well as new services to support citizens during the pandemic, could only be provided online. This has led many people who would never have approached digital technology before to transform their habits, and the city itself to set up services and plan projects tailored to citizens' needs.

It assumes that governance has been improved over time by receiving new features by coming up with smart governance which is the way government works with the participation of various stakeholders and the use of technology, bringing better citizen participation, and public and private partnerships. Accountability, cost reduction, linkages between the spheres of power, government-directed efforts, and innovation in public service provide higher quality public service delivery and consequent quality of life.

As a final outcome, smart governance aims to tie together all dimensions of the smart city and allows smart city projects to be undertaken. With smart governance, cities can improve transparency, reduce corruption, increase citizen participation, and promote partnerships between the public and private sectors. By leveraging ICT, smart governance helps enable smarter, more innovative, and more responsive cities.

### 3.1.2 Smart Environment

The concept of the smart environment is an essential component of smart city projects, as it encompasses the efficient management of energy and natural resources for achieving a better quality of life and a cleaner and more efficient environment. According to Vanolo (2014), the smart environment is understood in terms of the attractiveness of natural conditions, lack of pollution, and sustainable management of resources, suggesting that the smart environment is particularly related to the efficient management of energy and natural resources in the urban area.

Today, the concept of smart environment is increasingly gaining importance in the current society that is facing significant challenges such as overpopulation, lack of resources, and climate change. The increasing urbanization around the world is putting enormous pressure on natural resources, which is leading to environmental degradation, pollution, and depletion of natural resources (Chu, Cheng, & Yu, 2021). To counter these problems, smart cities are adopting practices that focus on the efficient management of energy and natural resources.

The smart environment dimension is critical in addressing these challenges because it promotes sustainable management of resources, attractive natural conditions, and reduced pollution levels. By adopting smart environment practices, cities can reduce their carbon footprint, improve their energy efficiency, and create a better quality of life for their citizens (Abid, Marchesani, Ceci, Masciarelli, & Ahmad, 2022; Colding & Barthel, 2017; Rohracher & Späth, 2014). These practices include the use of renewable energy sources, green buildings, efficient management of waste, and water resources, among others. For example, an important case already mentioned previously is that of the Nearly Zero-Emission Building (NZEB) concerning renewable energies and new infrastructures. In this regard, managers responsible for the allocation of funds have emphasized in interviews that:

> Currently, with the PNRR, the European Union requires us to make investments that bring a gain in terms of energy efficiency. It is important to note that this gain means not just improvement, as public works financed with smart city projects must bring an energy surplus. Therefore, these works should not only have zero impact but also generate additional clean energy that can be used by the city in its ordinary management. This approach is now a standard practice, and as an investment accountability office of the PNRR, we have the responsibility to continuously monitor and discuss these objectives, which, I repeat, are mandatory in the current city development plan through European funds for smart cities.

We assume that the efficient management of resources and the reduction of pollution levels will be central to the future development of cities, and that the direction taken by governments is to push and guide this transformation. This is because it is not only critical for the environment but also for the economic and social well-being of the citizens. The smart environment approach is, therefore, crucial in helping cities to achieve their sustainability goals and to address the challenges facing the world today, and the adoption of smart environmental practices is a significant step toward building sustainable and resilient cities that can withstand the challenges of the 21st century (Abid et al., 2022; Vanolo, 2014). For example, in Italy, environmental issues like air pollution, waste, and water management as well as climate change are rising in cities. The government has taken measures at the national level to mitigate these issues and reduce emissions but at the same time, environmental sustainability is also crucial part of the Sustainable Development Goals (SDGs) that Italy aims to achieve (Vanolo, 2014).

Although Italy's economy and society have developed unevenly across regions over the past decades, substantial differences remain in the progress of environmental and social development across cities (Leonelli, Marchesani, & Masciarelli, 2022). Thus, many Italian cities still face more traditional problems like lack of public services or deterioration of historical centers; they are redefining these issues in line with the smart city concept. Consequently, cities are reorganizing their agendas to transform into smart cities.

Cities shape the urban environment through infrastructures and buildings to enable future properties and meet the current migration from the urban areas to the cities. However, urban development consumes natural resources and energy, emits pollutants, and generates waste. Cities currently account for about 75% of the world's energy use and 70% of global $CO_2$ emissions, and these figures are rising as cities grow larger (Nasir, Duc Huynh, & Xuan Tram, 2019). The increasing intensity of urban metabolism and climate change are major sustainability challenges for cities, and the implementation of a smart environment dimension contributes to dealing with these issues.

Specifically, two main approaches to achieving a sustainable urban environment are considered. The first approach focuses on energy efficiency, renewable energy, green technologies, pollution control, and green management to reduce environmental impacts (Arabindoo, 2020; Gil-Garcia, Zhang, & Puron-Cid, 2016). The second approach centers on the efficient management of urban resources and utilities such as waste, water, transportation, and drainage to minimize pollution and improve resource quality (March & Ribera-Fumaz, 2016; Tachizawa, Alvarez-Gil, & Montes-Sancho, 2015). In this sense, smart mobility initiatives like reducing private vehicles and integrating transport also help decrease emissions and improve the environment as we can see in the next session.

Thus, it assumes that a smart environment is a key part of smart cities and aims to efficiently manage natural resources through technology and policies, and, in this sense, smart governance responsiveness to this challenge plays a central role in reacting to environmental contemporary challenges.

### 3.1.3 Smart Mobility

The need for improvements in urban infrastructure and subsystems has become a critical issue due to the rapid growth of urban populations and the high demand for quality of life. The increased demand for mobility has resulted in problems such as congestion and pollution (Cocchia, 2014). Following those issues, smart mobility has the potential to transform contemporary cities for the better as sustainable, integrated, and safe modes of transport must be considered in contemporary transport planning (Batty et al., 2012). Thus, smart mobility practices can offer users more transport options and more adaptable and affordable travel while reducing the reliance on private vehicles and promoting energy-efficient mobility (Benevolo, Dameri, & Auria, 2016).

In defining smart mobile, we must refer to a comprehensive concept that makes the transport network's sustainability more achievable due to the search for improvements in transport services, balancing the application of technology with social, economic, and environmental aspects (Vanolo, 2014). It is the ability to access transportation services from integrated platforms that aggregate the community and present intense processing of data from users to match the demand forecast.

Smart mobility is related to transport and the use of communication and information technologies to promote accessibility and increase the quality of life, and it includes the adoption of digital technologies that make mobility services in a city or territory more accessible and easier (Bakici, Almirall, & Wareham, 2013).

In implementing smart mobility practices, governments must be proactive and invest in sustainable transport solutions that promote energy saving. However, critical questions must be asked in terms of how the transition to smart mobility will be managed and how the benefits and any negative externalities of change will be governed (Docherty, Marsden, & Anable, 2018). The management of mobility by public and private agencies requires the adoption of intelligent and dynamic decision-making approaches capable of controlling various issues that change in real-time. Decision-making by responsible bodies must be linked to information such as the amount of data available, intelligent services, and energy results. They must also stick to issues that are specific to the reality of each city to implement relevant solutions that make mobility more intelligent (Vázquez, Lanero, Gutiérrez, & Sahelices, 2018). For example, a particularly noteworthy aspect of the interview discusses the opportunities and advantages of these practices.

> Our city policies are gradually adapting to the current mobility challenge and opportunities. The traditional means of transportation are changing, and cities are becoming smarter through the use of sensors, sustainable services, bike-sharing, and intelligent traffic lights, among others. For example, our city has a control-room that allows us to monitor the access to various public buildings, the movement of people on the streets, and in the city center, as well as in different areas. We work closely with the local police to ensure the safety of citizens and improve road safety.
>
> In addition to this, the control-room also helps us to monitor the vehicles passing through our city. We can check whether the vehicles are in compliance with the rules and regulations or not. We can also monitor the traffic flow and pollution levels (both noise pollution and air pollution) in real-time. This helps us to take necessary actions to control and reduce the pollution levels in our city.
>
> Moreover, the control-room helps us to monitor the speed of vehicles passing through the sensitive areas near schools. This allows us to take necessary actions to ensure the safety of children and improve road safety. By monitoring the traffic in real-time, we can offer better and more efficient services to the citizens of our city.

Thus, the urban mobility plan is a public management instrument that guides short-, medium-, and long-term projects, investments, and actions. It translates the objectives of improving local urban mobility into goals and strategic actions. Smart mobility requires a participatory process for preparing the plan, in which stakeholders identify challenges and propose improvement actions. They must also correlate the evolution in the usage dynamics of urban spaces with the transport system and consider the effect of new interventions on the natural and built environment and citizens' lives (Batty et al., 2012; Benevolo, Dameri, & D'Auria, 2016).

Despite internal organization and structural advancement, it is interesting to note that smart mobility is not only a means to solve internal problems, but it also offers a competitive advantage for the city when used strategically. Below is a practical example of this strategical perspective highlighted by a study conducted by Jan Henrik Nilsson (2019) on the city of Copenhagen:

> **Lessons from the City**: Copenhagen, Denmark
>
> The city of Copenhagen has leveraged the concept of smart mobility to enhance its transportation and attract new tourism, creating economic value for the entire area while simultaneously addressing mobility challenges. Specifically, cycling has been identified as a key aspect of smart mobility and urban planning efforts have been made to promote this mode of transportation. This includes the development of physical infrastructure such as bike lanes, commuting trails, intermodal solutions, bike-share schemes, and internet-based smart mobility.
>
> It is worth noting that bicycle tourism cannot be viewed separately from everyday cycling practices carried out by the local population. While the regional bicycle culture is highly path-dependent, urban planning strategies and the development of bicycle infrastructure form the basic conditions for urban bicycle tourism to flourish. This opens up new opportunities for visitors' interactions with the city and the local community.
>
> The local bicycle culture also creates awareness among tourists about possible cycling activities, which reinforces cycling as part of the destination identity. It becomes a unique aspect of the local experience that visitors are expected to participate in, contributing to a specific form of localness and generation an economic value for the city itself.

It assumes that smart mobility is an essential area of development for smart cities, as it impacts not only the mobility and internal management of a city but also its competitiveness and attractiveness. Smart mobility can be utilized as an economic and competitive engine for a city, generating value and contributing to its overall growth and development. Governments must be proactive in investing in sustainable transport solutions that promote energy-saving while ensuring that the transition to smart mobility is well-managed, and the benefits and any negative externalities of change are governed.

### 3.1.4 Smart Economy

The concept of a "smart economy" has emerged as a key aspect of the smart city paradigm. It embodies innovation, entrepreneurialism, labor market flexibility, integration into the international market, and the ability to transform business dynamics from a smart perspective. This section explores the smart economy dimension of smart cities, emphasizing its importance and elaborating on the factors that contribute to its development (Albino et al., 2015; Vanolo, 2014). It also explores how a well-developed smart economy creates an ecosystem that attracts and retains various stakeholders who interact with the city.

The smart economy is closely tied to a city's smart ecosystem and infrastructure, as well as the abundance of social and human capital. Together, these factors enable smart cities to develop more competitive business environments, attracting innovative companies, talent, students, knowledge, workforce, and investment (Vinod Kumar, 2020).

Consequently, smart cities contribute to establishing technology hubs that facilitate knowledge sharing through research centers, start-up incubators, accelerators, and innovation parks (Batty et al., 2012; Marchesani, Masciarelli, & Quang, 2022). Innovative cities and technology parks serve as natural magnets for establishing and promoting open innovation projects (Yun & Lee, 2019). For example, Morgan and Webb (2020) argue that when advanced IT infrastructures are developed locally through public–private partnerships, communities of lead users emerge in companies, public management, and university labs.

Moreover, the adoption of new technologies can rejuvenate markets and create economic benefits. By hosting firms that create such technologies, cities can gain economic superiority and competitiveness, attracting important company headquarters to the most developed cities (Camboim, Zawislak, & Pufal, 2019; Werner, 2002). This enhances the economic performance of cities both internally and externally when the technologies are effectively adopted and exported to other cities.

The prospect of shaping organizations according to European standards in terms of sustainability and innovation can potentially enhance their efficiency and help first to work in complex urban development. The ICT infrastructure of smart cities can also foster emerging and innovation-based business models, provided through regional, national, or multinational partnerships (Caragliu & Del Bo, 2019; Glaeser, 2000; Yigitcanlar, Kankanamge, & Vella, 2020).

However, public managers have highlighted the difficulty in matching public demand with private innovation, which brings both challenges and benefits to local firms. This difficulty is because the innovation brought by new technologies can be both challenging and beneficial, and local businesses may not always be able to respond promptly to these changes.

Despite the potential benefits of new technologies, dealing with contemporary technological innovations and meeting government requirements in terms of sustainability and digital reporting remains a challenge for local and national cities, as discussed in the previous section. This perspective is also evident in discussions with public managers who highlight the difficulty in creating a match between public demand and the level of innovation in the private sector:

> To date, as a city, we face significant challenges in engaging with local companies to implement these projects. There are very few companies that meet the European Union's requirements, and most of the time, we have to turn to foreign companies, mainly from the Netherlands, Poland, and Belgium. This is a paradox because there are many local companies that could collaborate and generate economic and social impact in the area. Unfortunately,

> their level of digitization holds them back, which also penalizes us as a city, resulting in longer lead times and a waste of resources.
>
> In the not-too-distant future, even local companies will have to adapt to this change because without these parameters, it will become increasingly difficult to participate in tenders and work with the public. In this case, the smart city promotes and encourages this transformation because if it doesn't, local companies will be severely penalized strategically. Therefore, they need to digitize and acquire the characteristics of sustainability and innovation, as in the rest of Europe.

Thus, despite the potential benefits of new technologies, dealing with contemporary technological innovations and meeting government requirements in terms of sustainability and digital reporting remains a challenge for local and national cities, as discussed in the previous section. However, the smart city is a driving force for this change, as it encourages local and national firms to adapt to these standards. Consequently, it is a challenge for local and national firms to shape their organizations according to European standards in terms of sustainability and innovation. This can potentially enhance their efficiency and help them work effectively in complex urban development.

In conclusion, it assumes that through the smart economy dimension, cities aim to become innovation-based, capital-based, and innovation hubs for businesses and human capital alike. Economic vibrancy embedded in the smart economy dimension is a key factor to affect a city's attractiveness and boosts the economic environment. When the smart economy dimension is well-developed, the city creates an ecosystem that draws in and retains all stakeholders who interact with it. By fostering innovation, collaboration, and the adoption of advanced technologies, smart cities can not only improve the well-being of their citizens but also create a sustainable, competitive, and thriving environment for businesses and talent to flourish.

### 3.1.5 Smart Living

Smart living is a crucial dimension in the development of smart cities. It encompasses initiatives that aim to improve health, education, social services, citizen participation, housing, and integration (Qonita & Giyarsih, 2022). However, the smart living dimension lacks clear conceptualization in the academic debate and presents challenges in terms of evaluating its impact on the territory (Vanolo, 2014).

The smart living dimension is not limited to the use of smart technology but emphasizes the quality of living produced by the independent implementation of smart technology under sustainable conditions. It aims to offer an urban context suited to the needs of its stakeholders by engaging with residents through advanced alternative or novel notions of urban and economic development, social inclusion, or greater urban planning (Camboim et al., 2019; Krishnan, Arumugam, & Maddulety, 2020; Richard Florida, 2002b). Citizens and users become the main actors in the smart transformation by becoming part of the remodeling and construction of the city itself. However, the relationship between citizens and the city is not always straightforward and presents social and environmental complexities. It is the city's responsibility to respond to and reduce this gap. One example that can relate to smart living practices is the city's support for its citizens during the pandemic, as reported in the following interview:

> During the pandemic, smart practices have become essential out of necessity. The city had to quickly understand the needs of its citizens and take drastic action to provide basic services for survival. This included not only education and healthcare services but also economic support to ensure that people could sustain themselves during difficult times. To address this challenge, we worked together with the government to provide meal vouchers and economic support in a fast and controlled manner that avoided people leaving their homes and creating crowds during the pandemic. We utilized the fiscal code card to distribute the support, and people in need adapted to this change, even those who were initially hesitant to use electronic money due to age or social construct.
>
> The city's response was prompt and effective, thanks to the use of current digital technologies. We were able to provide the necessary support to citizens and ensure their survival during the pandemic. …It is clear that smart practices are crucial for responding to the needs of citizens in times of crisis, and we will continue to utilize innovative solutions to improve the quality of life for our citizens.

As cities continue to transform into smart cities, it is essential for the smart living dimension to adapt and follow suit. This is especially important as migration flows continue to move toward major cities, increasing the need for improved health services, education, security, and integration in the urban environment to enhance the quality of life (B. N. Silva, Khan, & Han, 2018a; Xie, Zhou, & Luo, 2016).

One of the most significant challenges that arise with an increase in population is the strain placed on healthcare systems. As the demand for healthcare services increases, hospitals and healthcare facilities need to adapt to meet the growing needs of the population (Feng & Yuan, 2023). For instance, cities like Singapore have implemented smart health-care solutions such as telemedicine, home monitoring, and mobile healthcare apps to provide convenient and accessible health-care services to their citizens (Sharifi, Khavarian-Garmsir, & Kummitha, 2021). These solutions not only improve the quality of health care but also reduce the burden on physical healthcare facilities.

Similarly, the education system must also adapt to accommodate the growing population. As cities become more crowded, schools must find ways to accommodate more students while maintaining a high standard of education. One solution is to implement digital technologies to enhance the learning experience such as virtual reality (VR), gamification, and online learning platforms. Cities like Helsinki, Finland, have embraced digital technology in their education systems, providing free access to online learning platforms and digital tools to all students (Hämäläinen, 2020).

Security and integration are also crucial factors that contribute to the livability of urban environments. As cities grow, so does the diversity of its population, making it essential to implement initiatives that promote social inclusion and integration. For example, the city of Amsterdam has implemented a program called "We Are Here" that focuses on providing shelter and support to undocumented migrants in the city, helping to integrate them into society and reduce social inequality (Hajer & Bröer, 2020).

In the current debate around smart cities, dimensions such as the environment, mobility, and governance clearly impact the current local and economic environment. However, the dimension of smart living is increasingly crucial for residents and the community as it includes a social and inclusive perspective of the city that directly impacts not only service provision and improvement but also living conditions (Adler & Florida, 2021; S. Y. Tan & Taeihagh, 2020).

Thus, it assumes that the smart living dimension plays an essential role in the balance of a city and is indispensable in supporting citizens. Its initiatives make a difference in the context of cities, and it is necessary to integrate them into urban planning and development strategies. Today, the smart living dimension must adapt and follow the transformation of cities into smart cities.

### 3.1.6 Smart People

In the smart city dimensions, the last (but not least) aspect that is often overlooked in the smart city discourse is the role of the smart people dimension. According to the definition presented by Vanolo (2014), smart people dimension is linked to the level of qualification of human and social capital, flexibility, creativity, tolerance, cosmopolitanism, and participation in public life. In other words, smart people are individuals who possess a wide range of skills and competencies, are open-minded, and actively engage in their community. The aim of contemporary cities is to embrace and promote these characteristics.

Smart cities can foster both human capital and social capital development. Since human capital considers the skills and competencies embedded in an individual or a group, smart city projects aim to attract and offer a social and inclusive environment tailored to the user. For example, the growth in a metropolitan area's concentration of college-educated residents is not only directly correlated with employment growth but also to other factors such as inclusiveness, services, nightlife, opportunity, culture, and lifestyle. Not surprisingly, the ability to meet people and make friends is one of the most important factors that determine our happiness in our lives and communities (Batabyal & Nijkamp, 2019; Richard Florida, 2002a, 2002b). Human beings crave interaction, but, as Putnam reminds us in the famous book *Bowling alone: The collapse and revival of*

*American community*, human beings have an innate desire for social interaction, but modern society often isolates us and makes it difficult to find satisfying social connections and support (Putnam, 2000). Smart cities can contribute to improving this aspect of contemporary urban and social life.

Thus, as anticipated, smart cities consist of various dimensions, such as smart mobility, smart environment, smart economy, and smart governance. However, smart people also play a central role interacting with each of these dimensions.

In fact, everything that relates to people's lives with passion, enthusiasm, and interest should be promoted by the city. Creating a lively and virtuous environment contributes to increasing the attractiveness of the city and the positive perception of its citizens and users (Richard Florida, 2012). A city that only focuses on the economy or governance risks losing touch with the public, which is both a user and a customer and must be attracted and pampered by the city.

Based on these perspectives, we can take the example of Florence and Barcelona. These two cities are known for their rich cultural heritage and innovative use of technology to promote it. Florence has numerous museums, galleries, and historic sites that attract millions of tourists every year. The city has also developed smart solutions to enhance the visitor experience, such as a mobile app that provides information and guidance on the city's cultural offerings (Nevola, Coles, & Mosconi, 2022). Similarly, Barcelona has a vibrant cultural scene that includes music festivals, art exhibitions, and street performances. The city has also implemented smart solutions to promote its cultural heritage, such as a digital platform that provides access to thousands of images and videos of the city's historic and artistic treasures (Bourliataux-Lajoinie, Dosquet, & del Olmo Arriaga, 2019).

London is another example of a city that has successfully attracted people and developed businesses by focusing on its attractions, innovation, and technology. The city is known for its iconic landmarks, such as Big Ben and the Tower of London, as well as for its innovative start-ups and tech companies combining and integrating those innovative advancements into smart city projects (Taylor Buck & While, 2017).

In conclusion, smart cities need to invest in the smart people dimension to create vibrant and attractive environments that benefit both citizens and users. By promoting a lively and virtuous environment that embraces people's passions and interests, cities can increase their attractiveness, foster innovation, and create new business opportunities. Therefore, it is important for cities to consider this dimension to drive human and social capital, attract and retain talented individuals, and provide them with the social and inclusive environment that they need.

## 3.2 Smart City Dimensions Empower City Development

In the previous chapter, we discussed the concept of smart cities and explored the various dimensions that constitute a smart city ecosystem. These dimensions encompass economic growth, mobility, sustainability, human capital, living standards, and effective governance. In this chapter, we delve deeper into the internal and organizational perspective of smart city development, focusing on the orchestration and balance of these dimensions. By doing so, cities can maximize the potential benefits and overcome challenges associated with implementing a holistic smart city approach.

In this section, we will analyze the various dimensions that empower city development within the smart city framework. Each dimension plays a crucial role in enhancing economic growth, improving mobility and connectivity, promoting sustainability, fostering human capital and well-being, and ensuring effective governance.

First, the smart economy dimension plays a pivotal role in empowering economic growth within a smart city. By harnessing technology and data-driven approaches, cities can create an environment conducive to innovation, entrepreneurship, and economic diversification. Smart economies facilitate the development of knowledge-based industries, attracting investments, fostering job creation, and enhancing productivity (Hollands, 2015; Kummitha, 2018; Marchesani et al., 2023). Through the integration of digital platforms, smart cities can stimulate collaboration among businesses, academia, and research institutions, promoting knowledge transfer and creating a thriving ecosystem for economic development (Langley & Leyshon, 2017; Sancino & Hudson, 2020). Second, the smart mobility dimension focuses on improving transportation systems, enhancing connectivity, and ensuring efficient movement within the city. By integrating smart technologies such as intelligent transportation systems, real-time data analytics, and smart infrastructure, cities can optimize traffic flow, reduce congestion, and enhance transportation efficiency. Smart mobility solutions, such as shared mobility services, electric vehicles, and smart parking systems, contribute to reduced carbon emissions, improved air quality, and increased accessibility (Benevolo, Dameri, & Auria, 2016; Benevolo, Dameri, & D'Auria, 2016). Additionally, smart mobility initiatives can enhance connectivity between different modes of transportation, promoting seamless and sustainable travel experiences for residents and visitors (Seyfi & Hall, 2020; Vázquez et al., 2018). Third, the smart environment dimension emphasizes the integration of sustainable practices and technologies to preserve and enhance the natural environment. Smart cities prioritize resource efficiency, waste management, and renewable energy solutions to minimize environmental impact. By leveraging technologies like smart grids, energy management systems, and sensor networks, cities can optimize energy consumption, reduce greenhouse gas emissions, and promote sustainable living. Furthermore, smart environmental initiatives involve water management strategies, green spaces, and urban planning approaches that prioritize ecological balance and resilience (Abid et al., 2022; Arabindoo, 2020; Linde et al., 2021). Fourth, the smart people and smart living dimensions focus on enhancing the quality of life, well-being, and social inclusivity within a smart city. Smart cities recognize the importance of human capital development, access to education, and the provision of social services. By leveraging digital technologies and data-driven approaches, cities can offer personalized and accessible services, such as e-learning platforms, health-care innovations, and citizen engagement platforms. Smart living initiatives also involve the creation of sustainable and inclusive communities, with a focus on affordable housing, public spaces, and cultural amenities that enhance residents' well-being and social cohesion (Aletà, Alonso, & Ruiz, 2017; S. Kumar, Mookerjee, & Shubham, 2018; Linde et al., 2021; Vázquez et al., 2018). Finally, the smart governance dimension emphasizes the adoption of data-driven decision-making processes, citizen participation, and transparent and efficient governance structures. Smart cities leverage technology to improve service delivery, enhance public safety, and streamline administrative processes (Meijer & Bolívar, 2016; Pittaway & Montazemi, 2020). By implementing smart governance practices, cities can leverage data analytics, artificial intelligence, and digital platforms to enable evidence-based policymaking, proactive urban management, and effective resource allocation. Smart governance also promotes transparency and accountability, fostering trust and engagement between the government and the citizens (De Guimarães, Severo, Felix Júnior, Da Costa, & Salmoria, 2020; Lnenicka et al., 2022).

By understanding the interconnections and interdependencies among these dimensions, cities can orchestrate and balance them effectively. The success of a smart city lies in the ability to harmonize economic growth with sustainable mobility, environmental stewardship, human well-being, and efficient governance. Achieving this balance requires integrated planning, collaborative partnerships, and data-driven approaches (Linde et al., 2021; Tiwana, 2014; Vanolo, 2014).

In the next chapter, titled "Orchestrating and Balancing Smart City Dimensions," we will delve deeper into the strategies and best practices for achieving this harmony. We will explore how cities can integrate different dimensions, leverage technology and data, and engage stakeholders to create a cohesive smart city ecosystem. Through practical examples and case studies, we will highlight successful models of dimension orchestration and provide insights for future smart city projects.

As cities continue to evolve and face new challenges, the orchestration and balance of smart city dimensions become increasingly critical. By embracing a holistic approach, cities can create intelligent, sustainable, and inclusive urban environments that enhance the overall quality of life for their residents. The next chapter aims to provide an overview for achieving this vision, setting the stage for the implementation of effective strategies and the realization of the full potential of smart cities.

## 3.3 Orchestrating and Balancing Smart City Dimensions

Within this chapter, we explored the various dimensions that constitute a smart city ecosystem and their significance in empowering city development. However, it is essential to recognize that these dimensions are not isolated entities but rather interconnected and interdependent. To achieve the full potential of a smart city, it is crucial to adopt a holistic approach that orchestrates and balances these dimensions effectively.

The interconnections between smart city dimensions create a synergistic effect, where advancements in one dimension can enhance and support progress in others. For example, the development of a smart economy can drive job creation and economic growth, which, in turn, can improve the quality of life for residents and promote sustainable living (Linde et al., 2021; Tiwana, 2014; Vanolo, 2014). Likewise, investing in smart mobility can enhance connectivity and accessibility, leading to increased economic opportunities and improved social inclusivity. To orchestrate and balance smart city dimensions, cities need to adopt strategic approaches that facilitate collaboration, integration, and optimization. Here are three key strategies for achieving effective orchestration and balance.

### 3.3.1 Integrated Planning and Policy Frameworks

To ensure effective orchestration and balance of smart city dimensions, cities should develop comprehensive and integrated planning frameworks that consider all dimensions simultaneously (Camboim et al., 2019; Gil-Garcia et al., 2016). This approach allows cities to align the objectives and actions of each dimension, creating a cohesive and synergistic approach to smart city development. By integrating smart city goals into urban planning processes, cities can optimize resource allocation and minimize redundancies. For example, when planning infrastructure projects, cities can consider how the smart mobility dimension can be incorporated to enhance transportation systems, while also considering the impact on the smart environment dimension by incorporating sustainable design principles. Integrated planning frameworks enable cities to identify potential synergies and trade-offs between dimensions, ensuring that actions taken in one dimension positively contribute to others. For instance, a policy aimed at promoting economic growth through innovation and entrepreneurship can also have positive effects on the smart people dimension by creating job opportunities and fostering human capital development (Bibri & Krogstie, 2020; Fortino, Russo, Savaglio, Shen, & Zhou, 2018; Matos, Vairinhos, Dameri, & Durst, 2017).

> **Lesson from the city:** San Francisco, United States
>
> An exemplary city that has successfully implemented integrated planning and policy frameworks to drive its smart city initiatives is San Francisco, located in the United States. The city's approach has been centered around key dimensions such as transportation, sustainability, and digital innovation, showcasing the benefits of a cohesive and synergistic strategy.
> (J. H. Lee, Hancock, & Hu, 2014; Yigitcanlar, Han, Kamruzzaman, Ioppolo, & Sabatini-Marques, 2019)
>
> In this line, in the realm of transportation, San Francisco has adopted forward-thinking policies to promote the use of electric vehicles (EVs) and enhance public transportation infrastructure. By establishing charging infrastructure and incentivizing EV adoption, the city has successfully reduced carbon emissions and encouraged sustainable transportation alternatives. Additionally, San Francisco has integrated smart technologies and data-driven solutions into its public transportation systems, offering real-time transit information, intelligent traffic management, and smart parking solutions to improve mobility and alleviate congestion.

San Francisco's commitment to sustainability is evident through its policies and programs aimed at fostering green building development and utilizing renewable energy sources. By incorporating energy efficiency measures, water conservation initiatives, and effective waste management practices, the city has created a more sustainable urban environment. These efforts not only reduce the ecological footprint but also contribute to an improved quality of life for its residents.

(J. H. Lee et al., 2014)

Embracing digital innovation, San Francisco has harnessed the power of technology and data to enhance service delivery, optimize efficiency, and encourage innovation. The city has established open data platforms, employed smart city analytics, and implemented digital engagement platforms to promote citizen participation. By leveraging these digital tools and fostering innovation ecosystems, San Francisco has created an inclusive and connected city that empowers its residents and businesses.

(Adler & Florida, 2021; J. H. Lee et al., 2014; Yigitcanlar et al., 2019)

San Francisco's success in integrating planning and policy frameworks across multiple dimensions showcases the importance of aligning goals and actions. By considering the interdependencies between transportation, sustainability, and digital innovation, the city has optimized resource allocation and minimized redundancies. San Francisco's comprehensive planning framework serves as a practical example of how cities can orchestrate and balance smart city dimensions to create a sustainable and inclusive urban environment that leverages technology for the benefit of its residents.

### 3.3.2 Collaborative Partnerships and Stakeholder Engagement

Effective orchestration of smart city dimensions necessitates collaboration and engagement among various stakeholders. This includes government agencies, businesses, community organizations, academia, and citizens. Collaborative partnerships enable cities to leverage diverse expertise, resources, and perspectives to address complex challenges and develop innovative solutions. Engaging stakeholders in the decision-making process promotes inclusivity, ownership, and a shared vision for the city's future. By involving citizens, for example, in the cocreation of smart city initiatives, cities can ensure that solutions are tailored to the specific needs and aspirations of the community. Collaboration with businesses and academia can bring technical expertise and research capabilities, driving innovation and implementation of cutting-edge technologies (Abid et al., 2022; Lei, Ye, Wang, & Law, 2020; Zhao, Fashola, Olarewaju, & Onwumere, 2021).

Collaborative partnerships also facilitate knowledge sharing and learning among stakeholders. By exchanging best practices and lessons learned, cities can accelerate their smart city development journey and avoid potential pitfalls. Furthermore, partnerships with community organizations can ensure that the social dimensions of a smart city, such as equity and inclusivity, are adequately addressed in the planning and implementation processes.

> **Lesson from the city:** Lisbon, Portugal
>
> Lisbon has been proactive in fostering collaboration and engaging stakeholders in its smart city development. The city has established the Lisbon Urban Data Laboratory, which serves as a collaborative platform for various stakeholders, including local government, businesses, and academia. This platform enables these entities to work together in co-creating and implementing smart city solutions.[1]
>
> Through the Lisbon Urban Data Laboratory, stakeholders collaborate on projects focused on key smart city areas such as energy, waste management, mobility, and social inclusion. By involving diverse stakeholders, Lisbon ensures that multiple perspectives and expertise are considered, leading to more comprehensive and effective solutions. This collaborative approach also helps to leverage resources, share knowledge, and avoid duplication of efforts.
>
> Moreover, Lisbon has actively engaged its citizens in the smart city development process. The city has implemented the Smart Lisbon platform, an online platform that encourages citizen participation and contribution. Through this platform, residents can provide their ideas, suggestions, and feedback on various smart city initiatives. This engagement empowers citizens to play an active role in shaping the future of their city and ensures that solutions are tailored to meet their specific needs and aspirations.[1]
>
> (Albuquerque et al., 2021)

---

[1] EPRS: European Parliamentary Research Service. Belgium. Retrieved from https://policycommons.net/artifacts/1339578/mapping-smart-cities-in-the-eu/1949353/

> The collaborative partnerships and stakeholder engagement in Lisbon's smart city initiatives have yielded significant benefits. By involving businesses, academia, and citizens, the city taps into a wealth of expertise and resources, enabling the implementation of cutting-edge technologies and innovative solutions. The active engagement of citizens fosters a sense of ownership and inclusivity, ensuring that the smart city initiatives align with the priorities and values of the community. Lisbon's collaborative approach has resulted in tangible outcomes. The city has made remarkable progress in areas such as energy efficiency, mobility, and social inclusion. For example, initiatives like smart lighting systems, electric vehicle charging infrastructure, and community-driven programs have been successfully implemented. These efforts have not only improved the quality of life for residents but have also positioned Lisbon as a leading smart city on the global stage.
>
> (Camboim et al., 2019)

In this line, Lisbon's case exemplifies the importance of collaborative partnerships and stakeholder engagement in driving smart city initiatives. By involving diverse stakeholders and actively engaging citizens, Lisbon has been able to tap into collective knowledge, resources, and perspectives to develop innovative and sustainable solutions. The collaborative approach has not only led to successful implementation but has also fostered a sense of ownership and inclusivity, ensuring that the smart city transformation is driven by the needs and aspirations of the community.

### 3.3.3 Data Sharing and Interoperability

Data are a crucial resource for smart city operations and decision-making. Cities should prioritize data sharing and interoperability among different dimensions to enable comprehensive analysis and informed decision-making. Breaking down data silos and establishing open data platforms are essential steps in unlocking the full potential of data-driven approaches and facilitating cross-dimensional integration. By enabling data sharing, cities can combine data from various sources and dimensions to gain a holistic understanding of their urban challenges and opportunities. For instance, by integrating data from smart mobility systems with environmental sensors, cities can analyze the impact of transportation on air quality and identify strategies to reduce emissions (Dewi, Hidayanto, Purwandari, Kosandi, & Budi, 2018; Löfgren & Webster, 2020; Oztemel & Gursev, 2020; Pittaway & Montazemi, 2020). Interoperability is also critical in enabling the seamless exchange and integration of data between different systems and stakeholders. Standardized data formats, protocols, and interfaces allow for efficient data flow and analysis across dimensions. This interoperability empowers cities to generate valuable insights, improve efficiency, and enhance the overall performance of their smart city initiatives.

**Lesson from the city:** Barcelona, Spain

Barcelona, located in Spain, has emerged as a leading smart city by embracing data-driven approaches and prioritizing data sharing and interoperability among different dimensions of the city. The city has implemented the Sentilo platform, an open-source initiative that serves as a data-sharing hub for various smart city systems.[2]

One notable aspect of Barcelona's approach is its integration of data from sensors, devices, and applications across different dimensions, including transportation, energy, and environment. For example, the city has deployed smart sensors to collect real-time data on traffic patterns, air quality, and energy consumption. By combining data from these diverse sources, Barcelona gains a comprehensive understanding of urban challenges and opportunities. Through the Sentilo platform, Barcelona enables stakeholders to securely access and share data, fostering collaboration and innovation (Bakici et al., 2013; Bourliataux-Lajoinie et al., 2019). Government agencies, businesses, research institutions, and citizens can leverage the platform to develop data-driven solutions tailored to the city's needs. This collaborative approach ensures that data is harnessed effectively to address urban challenges and improve the quality of life for residents.

Barcelona's focus on data sharing and interoperability has facilitated cross-domain analysis, allowing the city to explore connections and synergies between different dimensions. For instance, by integrating data from transportation systems with environmental sensors, Barcelona can analyze the impact of mobility on air quality and develop strategies to reduce emissions. This interconnected approach helps optimize urban services, enhance sustainability, and create a more livable environment for its residents. Furthermore, Barcelona's emphasis on open-source initiatives like Sentilo fosters innovation and knowledge sharing (Sinaeepourfard et al., 2016). The platform enables developers and entrepreneurs to build applications and services based on shared data, driving economic growth and technological advancements. By encouraging a collaborative ecosystem, Barcelona leverages the expertise and creativity of a wide range of stakeholders to tackle urban challenges effectively (Bakici et al., 2013; Camboim et al., 2019; Nesti & Graziano, 2020; Sinaeepourfard et al., 2016).

The success of Barcelona's data sharing and interoperability initiatives can be attributed to the city's holistic approach to smart city development. By integrating data from various dimensions, engaging stakeholders, and fostering collaboration, Barcelona has created a thriving ecosystem that harnesses the power of data to transform urban life. Overall, Barcelona's case study exemplifies the importance of data sharing and interoperability in smart city initiatives. By breaking down data silos and establishing open platforms, cities can unlock the full potential of data-driven approaches, foster collaboration, and drive innovation for the benefit of their residents.

Drawing upon the aforementioned examples and in-depth discussions, it becomes evident that the orchestration and balance of smart city dimensions hold immense significance for cities aspiring to fully realize the potential of their smart city initiatives. By strategically integrating and aligning these dimensions, cities can foster a comprehensive and synergistic approach to urban development, thereby unlocking a myriad of opportunities and reaping substantial benefits.

Firstly, by orchestrating these dimensions, cities can leverage their unique characteristics and strengths to maximize the impact of their smart city initiatives. Every city has its distinct identity, economic landscape, natural resources, and societal makeup. By understanding and considering these intrinsic characteristics, cities can tailor their strategies and investments to align with their specific needs and goals (Coca-Stefaniak, 2021; Tiwana, 2014; Vanolo, 2014). For instance, a city known for its technological innovation can focus on developing a robust smart economy dimension, while a city with abundant green spaces can prioritize the smart environment dimension to enhance sustainability. Moreover, orchestrating smart city dimensions offers cities the opportunity to address interconnected challenges and seize synergies. For example, integrating the smart mobility dimension with the smart environment dimension can lead to solutions that reduce congestion, improve air quality, and promote sustainable transportation options. By recognizing and capitalizing on these interdependencies, cities can create comprehensive and integrated solutions that enhance the overall performance and resilience of their urban systems. Additionally, effective orchestration of smart city dimensions enables cities to optimize resource allocation and minimize redundancies. By taking a coordinated approach, cities can avoid fragmented and isolated initiatives, ensuring that investments and efforts are directed toward common objectives. This not only improves the efficiency and effectiveness of smart city projects but also maximizes the return on investment for both the city and its stakeholders. Finally, by successfully orchestrating these dimensions, cities can create a unique and attractive urban environment that fosters innovation, economic growth, and quality of life. A well-balanced and harmonious integration of smart city dimensions can enhance the city's competitiveness, attract talent, and stimulate economic activities. It can also improve the livability and well-being of residents by providing sustainable infrastructure, equitable access to services, and a vibrant community fabric.

In conclusion, orchestrating and balancing smart city dimensions is crucial for cities to unlock their full potential and capitalize on the opportunities presented by the smart city paradigm. By understanding their own characteristics and aligning their strategies with the dimensions that best represent and complement their city, urban centers can create a thriving ecosystem that benefits residents, businesses, and the overall urban environment. The effective orchestration of smart city dimensions will play a pivotal role in shaping the cities of the future, enabling sustainable, inclusive, and prosperous urban environments for generations to come.

## Notes

1. https://urbandatalab.pt/index.php/the-lab/outputs/lisbon-urban-data-labs
2. *Sentilo* website: https://ajuntament.barcelona.cat/digital/en/digital-transformation/urban-technology/sentilo and https://connecta.bcn.cat/

**References:**

Abid, N., Marchesani, F., Ceci, F., & Masciarelli, F. (2023). Assessing Capabilities to Embrace Digital Transformation: The Case of Southern Italy. In *Lecture Notes in Information Systems and Organisation* (Vol. 59, pp. 169–182). https://doi.org/10.1007/978-3-031-15770-7_11

Abid, N., Marchesani, F., Ceci, F., Masciarelli, F., & Ahmad, F. (2022). Cities trajectories in the digital era : Exploring the impact of technological advancement and institutional quality on environmental and social sustainability. *Journal of Cleaner Production*, *377*(September), 134378. https://doi.org/10.1016/j.jclepro.2022.134378

Adler, P., & Florida, R. (2021). The rise of urban tech: how innovations for cities come from cities. *Regional Studies*, *55*(10–11), 1787–1800. https://doi.org/10.1080/00343404.2021.1962520

Albino, V., Berardi, U., & Dangelico, R. M. (2015). Smart cities: Definitions, dimensions, performance, and initiatives. *Journal of Urban Technology*. https://doi.org/10.1080/10630732.2014.942092

Arabindoo, P. (2020). Renewable energy, sustainability paradox and the post-urban question. *Urban Studies*, *57*(11), 2300–2320. https://doi.org/10.1177/0042098019885080

Bakıcı, T., Almirall, E., & Wareham, J. (2013). A Smart City Initiative: The Case of Barcelona. *Journal of the Knowledge Economy*, *4*(2). https://doi.org/10.1007/s13132-012-0084-9

Batabyal, A. A., & Nijkamp, P. (2019). Creative capital, information and communication technologies, and economic growth in smart cities. *Economics of Innovation and New Technology*, *28*(2). https://doi.org/10.1080/10438599.2018.1433587

Batty, M., Axhausen, K. W., Giannotti, F., Pozdnoukhov, A., Bazzani, A., Wachowicz, M., Ouzounis, G., & Portugali, Y. (2012). Smart cities of the future. *European Physical Journal: Special Topics*, *214*(1), 481–518. https://doi.org/10.1140/epjst/e2012-01703-3

Benevolo, C., Dameri, R. P., & D'Auria, B. (2016). *Smart Mobility in Smart City*. https://doi.org/10.1007/978-3-319-23784-8_2

Borrás, S., & Edler, J. (2020). The roles of the state in the governance of socio-technical systems' transformation. *Research Policy*, *49*(5), 103971. https://doi.org/10.1016/j.respol.2020.103971

Bourliataux-Lajoinie, S., Dosquet, F., & del Olmo Arriaga, J. L. (2019). The dark side of digital technology to overtourism: the case of Barcelona. *Worldwide Hospitality and Tourism Themes*, *11*(5). https://doi.org/10.1108/WHATT-06-2019-0041

Camboim, G. F., Zawislak, P. A., & Pufal, N. A. (2019). Driving elements to make cities smarter: Evidences from European projects. *Technological Forecasting and Social Change*, *142*(December 2017), 154–167. https://doi.org/10.1016/j.techfore.2018.09.014

Caragliu, A., & Del Bo, C. F. (2019). Smart innovative cities: The impact of Smart City policies on urban innovation. *Technological Forecasting and Social Change*, *142*(December 2017), 373–383. https://doi.org/10.1016/j.techfore.2018.07.022

Carvalho, L. (2015). Smart cities from scratch? A socio-technical perspective. *Cambridge Journal of Regions, Economy and Society*, *8*(1), 43–60. https://doi.org/10.1093/cjres/rsu010

Chu, Z., Cheng, M., & Yu, N. N. (2021). A smart city is a less polluted city. *Technological Forecasting and Social Change*, *172*(April 2020), 121037. https://doi.org/10.1016/j.techfore.2021.121037

Cocchia, A. (2014). *Smart and Digital City: A Systematic Literature Review*. https://doi.org/10.1007/978-3-319-06160-3_2

Colding, J., & Barthel, S. (2017). An urban ecology critique on the "Smart City" model. *Journal of Cleaner Production*, *164*, 95–101. https://doi.org/10.1016/j.jclepro.2017.06.191

Dirks, S., & Keeling, M. (2009). A vision of smarter cities: How cities can lead the way into a prosperous and sustainable future. *IBM Institute for Business Value. June*.

Docherty, I., Marsden, G., & Anable, J. (2018). The governance of smart mobility. *Transportation Research Part A: Policy and Practice*. https://doi.org/10.1016/j.tra.2017.09.012

Feng, W., & Yuan, H. (2023). The impact of medical infrastructure on regional innovation: An empirical analysis of China's prefecture-level cities. *Technological Forecasting and Social Change*, *186*(PA), 122125. https://doi.org/10.1016/j.techfore.2022.122125

Florida, R. (2002a). The economic geography of talent. *Annals of the Association of American Geographers*, *92*(4), 743–755. https://doi.org/10.1111/1467-8306.00314

Florida, R. (2002b). The Rise of the Creative Class. *Washington Monthly*.

Florida, R. (2012). *The Rise of the Creative Class--Revisited: 10th Anniversary Edition--Revised and Expanded*. http://www.amazon.com/The-Rise-Creative-Class-Revisited-Edition-Revised/dp/0465029930

Gagliardi, D., Schina, L., Sarcinella, M. L., Mangialardi, G., Niglia, F., & Corallo, A. (2017). Information and communication technologies and public participation: interactive maps and value added for citizens. *Government Information Quarterly*, *34*(1), 153–166. https://doi.org/10.1016/j.giq.2016.09.002

Giffinger, R., Fertner, C., Kramar, H., & Meijers, E. (2007). City-ranking of European medium-sized cities. In *Centre of Regional Science, Vienna UT*.

Gil-Garcia, J. R., Zhang, J., & Puron-Cid, G. (2016). Conceptualizing smartness in government: An integrative and multi-dimensional view. *Government Information Quarterly*, *33*(3), 524–534. https://doi.org/10.1016/j.giq.2016.03.002

Glaeser, E. L. (2000). The New Economics of Urban and Regional Growth. In *The American Economic Review*.

Hajer, M., & Bröer, C. (2020). We Are Here! Claim-making and Claim-placing of Undocumented Migrants in Amsterdam. *European Journal of Cultural and Political Sociology*, *7*(4). https://doi.org/10.1080/23254823.2020.1774911

Hämäläinen, M. (2020). A Framework for a Smart City Design: Digital Transformation in the Helsinki Smart City. In *Contributions to Management Science*. https://doi.org/10.1007/978-3-030-23604-5_5

Hollands, R. G. (2008). Will the real smart city please stand up? Intelligent, progressive or entrepreneurial? *City*, *12*(3), 303–320. https://doi.org/10.1080/13604810802479126

Krishnan, B., Arumugam, S., & Maddulety, K. (2020). Critical success factors for the digitalization of smart cities. *International Journal of Technology Management and Sustainable Development*, *19*(1), 69–86. https://doi.org/10.1386/tmsd_00016_1

Kummitha, R. K. R. (2020). Why distance matters: The relatedness between technology development and its appropriation in smart cities. *Technological Forecasting and Social Change*, *157*(April), 120087. https://doi.org/10.1016/j.techfore.2020.120087

Lee, S. Y., Florida, R., & Gates, G. (2010). Innovation, human capital, and creativity. *International Review of Public Administration*. https://doi.org/10.1080/12294659.2010.10805158

Leonelli, S., Marchesani, F., & Masciarelli, F. (2022). Risk or Opportunity? Exploring the Relationship Between Entrepreneurial Decision and the Use of Equity Crowdfunding Campaigns in Less- and Well-Developed Regions in Ital. In *The International Dimension of Entrepreneurial Decision-Making* (pp. 99–114). https://doi.org/10.1016/S0140-6736(46)91888-0

Linde, L., Sjödin, D., Parida, V., & Wincent, J. (2021). Dynamic capabilities for ecosystem orchestration A capability-based framework for smart city innovation initiatives. *Technological Forecasting and Social Change*, *166*. https://doi.org/10.1016/j.techfore.2021.120614

March, H., & Ribera-Fumaz, R. (2016). Smart contradictions: The politics of making Barcelona a Self-sufficient city. *European Urban and Regional Studies*, *23*(4), 816–830. https://doi.org/10.1177/0969776414554488

Marchesani, F., & Ceci, F. (2023). *Digitalization , Online Services , and Entrepreneurial Environment in the Italian Smart Cities ' Transition*.

Marchesani, F., Iaia, L., Masciarelli, F., & Christofi, M. (2022). Smart City ' s Internationalization and International Management Strategies in the Digital Era : a Systematic Literature Review. *Sustainable Digital Transformation*, 153–165.

Marchesani, F., & Masciarelli, F. (2021). *Crowdfunding as Entrepreneurial Investment: The Role of Local Knowledge Spillover*. 92–108. https://doi.org/10.1007/978-3-030-87842-9_8

Marchesani, F., Masciarelli, F., & Bikfalvi, A. (2023a). Cities (r)evolution in the smart era: smart mobility practices as a driving force for tourism flow and the moderating role of airports in cities. *International*

*Journal of Tourism Cities*, *9*(4), 1025–1045. https://doi.org/10.1108/IJTC-05-2023-0104

Marchesani, F., Masciarelli, F., & Bikfalvi, A. (2023b). Smart city as a hub for talent and innovative companies: Exploring the (dis) advantages of digital technology implementation in cities. *Technological Forecasting and Social Change*, *193*(May), 122636. https://doi.org/10.1016/j.techfore.2023.122636

Marchesani, F., Masciarelli, F., & Doan, H. Q. (2022). Innovation in cities a driving force for knowledge flows: Exploring the relationship between high-tech firms, student mobility, and the role of youth entrepreneurship. *Cities*, *130*, 103852. https://doi.org/10.1016/j.cities.2022.103852

Marsal-Llacuna, M. L., Colomer-Llinàs, J., & Meléndez-Frigola, J. (2015). Lessons in urban monitoring taken from sustainable and livable cities to better address the Smart Cities initiative. *Technological Forecasting and Social Change*, *90*(PB), 611–622. https://doi.org/10.1016/j.techfore.2014.01.012

Marsal-Llacuna, M. L., & Segal, M. E. (2017). The Intelligenter Method (II) for "smarter" urban policy-making and regulation drafting. *Cities*, *61*. https://doi.org/10.1016/j.cities.2016.05.006

Mergel, I., Edelmann, N., & Haug, N. (2019). Defining digital transformation: Results from expert interviews. *Government Information Quarterly*, *36*(4). https://doi.org/10.1016/j.giq.2019.06.002

Morgan, K., & Webb, B. (2020). Googling the city: In search of the public interest on Toronto's 'smart' waterfront. *Urban Planning*, *5*(1). https://doi.org/10.17645/UP.V5I1.2520

Nam, T., & Pardo, T. A. (2011). Conceptualizing smart city with dimensions of technology, people, and institutions. *ACM International Conference Proceeding Series*. https://doi.org/10.1145/2037556.2037602

Nasir, M. A., Duc Huynh, T. L., & Xuan Tram, H. T. (2019). Role of financial development, economic growth & foreign direct investment in driving climate change: A case of emerging ASEAN. *Journal of Environmental Management*, *242*(January), 131–141. https://doi.org/10.1016/j.jenvman.2019.03.112

Neirotti, P., De Marco, A., Cagliano, A. C., Mangano, G., & Scorrano, F. (2014). Current trends in smart city initiatives: Some stylised facts. *Cities*. https://doi.org/10.1016/j.cities.2013.12.010

Neumann, O., Matt, C., Hitz-Gamper, B. S., Schmidthuber, L., & Stürmer, M. (2019). Joining forces for public value creation? Exploring collaborative innovation in smart city initiatives. *Government Information Quarterly*, *36*(4), 101411. https://doi.org/10.1016/j.giq.2019.101411

Nevola, F., Coles, T., & Mosconi, C. (2022). Hidden Florence revealed? Critical insights from the operation of an augmented reality app in a World Heritage City. *Journal of Heritage Tourism*, *17*(4). https://doi.org/10.1080/1743873X.2022.2036165

Nilsson, J. H. (2019). Urban bicycle tourism: path dependencies and innovation in Greater Copenhagen. *Journal of Sustainable Tourism*, *27*(11), 1648–1662. https://doi.org/10.1080/09669582.2019.1650749

Pinna, F., Masala, F., & Garau, C. (2017). Urban policies and mobility trends in Italian smart cities. *Sustainability (Switzerland)*, *9*(4). https://doi.org/10.3390/su9040494

Pittaway, J. J., & Montazemi, A. R. (2020). Know-how to lead digital transformation: The case of local governments. *Government Information Quarterly*. https://doi.org/10.1016/j.giq.2020.101474

Putnam, R. D. (2000). Bowling Alone: The Collapse and Revival of American Community: New York: Simon und Schuster, 2001. ISBN. *Policy Analysis*, *20*.

Qonita, M., & Giyarsih, S. R. (2022). Smart city assessment using the Boyd Cohen smart city wheel in Salatiga, Indonesia. *GeoJournal*, *0123456789*. https://doi.org/10.1007/s10708-022-10614-7

Rohracher, H., & Späth, P. (2014). The Interplay of Urban Energy Policy and Socio-technical Transitions: The Eco-cities of Graz and Freiburg in Retrospect. *Urban Studies*, *51*(7). https://doi.org/10.1177/0042098013500360

Sancino, A., & Hudson, L. (2020). Leadership in, of, and for smart cities–case studies from Europe, America, and Australia. *Public Management Review*, *22*(5), 701–725. https://doi.org/10.1080/14719037.2020.1718189

Scholl, H. J., & Alawadhi, S. (2016). Creating Smart Governance: The key to radical ICT overhaul at the City of Munich. *Information Polity*, *21*(1). https://doi.org/10.3233/IP-150369

Sharifi, A., Khavarian-Garmsir, A. R., & Kummitha, R. K. R. (2021). Contributions of smart city solutions and technologies to resilience against the covid-19 pandemic: A literature review. *Sustainability (Switzerland)*, *13*(14). https://doi.org/10.3390/su13148018

Shelton, T., Zook, M., & Wiig, A. (2015). The "actually existing smart city." *Cambridge Journal of Regions, Economy and Society*, *8*(1), 13–25. https://doi.org/10.1093/cjres/rsu026

Silva, B. N., Khan, M., & Han, K. (2018). Towards sustainable smart cities: A review of trends, architectures, components, and open challenges in smart cities. *Sustainable Cities and Society*, *38*(January), 697–713. https://doi.org/10.1016/j.scs.2018.01.053

Tachizawa, E. M., Alvarez-Gil, M. J., & Montes-Sancho, M. J. (2015). How "smart cities" will change supply chain management. *Supply Chain Management*, *20*(3), 237–248. https://doi.org/10.1108/SCM-03-2014-0108

Tan, S. Y., & Taeihagh, A. (2020). Smart city governance in developing countries: A systematic literature review. *Sustainability (Switzerland)*. https://doi.org/10.3390/su12030899

Taylor Buck, N., & While, A. (2017). Competitive urbanism and the limits to smart city innovation: The UK Future Cities initiative. *Urban Studies*, *54*(2), 501–519. https://doi.org/10.1177/0042098015597162

Tura, N., & Ojanen, V. (2022). Sustainability-oriented innovations in smart cities: A systematic review and emerging themes. *Cities*, *126*(April), 103716. https://doi.org/10.1016/j.cities.2022.103716

Vanolo, A. (2014). Smartmentality: The Smart City as Disciplinary Strategy. *Urban Studies*, *51*(5), 883–898. https://doi.org/10.1177/0042098013494427

Vázquez, J. L., Lanero, A., Gutiérrez, P., & Sahelices, C. (2018). *The Contribution of Smart Cities to Quality of Life from the View of Citizens*. 55–66. https://doi.org/10.1007/978-3-319-71014-3_3

Vinod Kumar, T. M. (2020). Smart environment for smart cities. In *Advances in 21st Century Human Settlements*. https://doi.org/10.1007/978-981-13-6822-6_1

Vu, K., & Hartley, K. (2018). Promoting smart cities in developing countries: Policy insights from Vietnam. *Telecommunications Policy*, *42*(10), 845–859. https://doi.org/10.1016/j.telpol.2017.10.005

Werner, S. (2002). Recent Developments in International Management Research: A Review of 20 Top Management Journals. In *Journal of Management*. https://doi.org/10.1177/014920630202800303

Xie, B., Zhou, J., & Luo, X. (2016). Mapping spatial variation of population aging in China's mega cities. *Journal of Maps*, *12*(1), 181–192. https://doi.org/10.1080/17445647.2014.1000984

Yigitcanlar, T., Kankanamge, N., & Vella, K. (2020). How Are Smart City Concepts and Technologies Perceived and Utilized? A Systematic Geo-Twitter Analysis of Smart Cities in Australia. *Journal of Urban Technology*, *0*(0), 1–20. https://doi.org/10.1080/10630732.2020.1753483

Yun, Y., & Lee, M. (2019). Smart City 4.0 from the perspective of open innovation. *Journal of Open Innovation: Technology, Market, and Complexity*, *5*(4). https://doi.org/10.3390/joitmc5040092